\documentclass[prl,longbibliography,aps,reprint,showpacs,raggedbottom,nofootinbib,superscriptaddress,floatfix]{revtex4-2}

\usepackage{graphicx}
\usepackage{dcolumn}
\usepackage{bm}
\usepackage{hyperref}
\usepackage{gensymb}
\usepackage{siunitx}
\usepackage{amsmath}
\usepackage{amsfonts}
\usepackage{xcolor}
\usepackage{array,multirow}

\newcommand{\ignore}[1]{}

\begin{document}

\title{Direct measurement of the lifetime and coherence time of $\mathbf{\mathrm{Cu_2O}}$ Rydberg excitons}

\author{Poulab Chakrabarti}
\thanks{These authors contributed equally to this work.}
\affiliation{Universit\'e de Toulouse, INSA-CNRS-UPS, LPCNO, 135 Av. Rangueil, 31077 Toulouse, France}
\author{Kerwan Morin}
\thanks{These authors contributed equally to this work.}
\affiliation{Universit\'e de Toulouse, INSA-CNRS-UPS, LPCNO, 135 Av. Rangueil, 31077 Toulouse, France}
\author{Delphine Lagarde}
\affiliation{Universit\'e de Toulouse, INSA-CNRS-UPS, LPCNO, 135 Av. Rangueil, 31077 Toulouse, France}
\author{Xavier Marie}
\affiliation{Universit\'e de Toulouse, INSA-CNRS-UPS, LPCNO, 135 Av. Rangueil, 31077 Toulouse, France}
\affiliation{Institut Universitaire de France, 75231 Paris, France}
\author{Thomas Boulier}
\email{boulier@insa-toulouse.fr}
\affiliation{Universit\'e de Toulouse, INSA-CNRS-UPS, LPCNO, 135 Av. Rangueil, 31077 Toulouse, France}

\date{\today}

\begin{abstract}
Rydberg states of excitons are promising quantum objects to engineer giant nonlinearities in a solid-state system. For this purpose, a deeper understanding of the dynamics of Rydberg excitons and of their potential for coherent manipulation becomes important. We report experimental results where two-photon absorption is resonant with various Rydberg states of excitons in copper oxide and we detect their emission dynamics on a streak camera with sub-picosecond resolution. This technique enables the direct measurement of the Rydberg states lifetimes, which are in good agreement with the expected Rydberg scaling law. Moreover, we observe several intriguing dynamics including the presence of long-lived coherent oscillations. Finally, we independently measure the coherence time of the Rydberg states using a modified Michelson interferometer and find a good agreement with the coherent oscillations detected in the exciton emission dynamics. The lifetimes also reveal the absence of inhomogeneous broadening in the current high-precision spectroscopic data for the S series, which together with the presence of significant coherence time confirms the suitability of the system for coherent engineering. 
\end{abstract}

\maketitle


Coherence is a major quantum resource, necessary for any quantum application. As such, it is important to accurately assess the (unavoidably finite) coherence time of real-life quantum systems. Moreover, many such systems rely on the manipulation of excited states, which possess a finite lifetime. This lifetime represents an additional source of decoherence and it is therefore crucial to fully characterize both the lifetimes and coherence times involved in emerging quantum platforms.

Rydberg states of excitons have garnered recent interest due to their unique quantum properties and potential applications in quantum technologies~\cite{heckotter2024rydberg,taylor2022simulation, assmann2020semiconductor}. Excitons, which are bound electron-hole pairs present in semiconductors, can exhibit Rydberg states when they are excited to high principal quantum numbers $n$~\cite{kazimierczuk2014giant}. This leads to significantly enlarged orbitals and enhanced interactions, both between excitons and with external fields~\cite{heckotter2017scaling,heckotter2017high,kruger2020interaction}. In cuprous oxide ($\mathrm{Cu_2O}$), a well-known direct bandgap semiconductor, Rydberg excitons manifest as a series of discrete energy levels akin to hydrogen atoms, with distinct modifications due to the crystal structure, dielectric screening, and exciton mass~\cite{assmann2020semiconductor}. Recent advances in spectroscopic techniques have enabled the observation of these states up to $n=30$~\cite{versteegh2021giant}, which corresponds to orbitals spanning several microns in diameter. This opens avenues for exploring fundamental quantum phenomena, such as strongly nonlinear light-matter interactions~\cite{walther2018giant,morin2022self,pritchett2024giant,makhonin2024nonlinear} and long-range dipole-dipole interactions~\cite{walther2018interactions,walther2020plasma}, which are promising for applications in quantum information processing and excitonic devices~\cite{adams2019rydberg,assmann2020semiconductor}.

Recent research on Rydberg excitons enabled a deeper understanding of their fundamental nature~\cite{kazimierczuk2014giant,heckotter2017scaling,rogers2022high}, their interactions with the crystal environment~\cite{stolz2018interaction,kruger2020interaction,heckotter2020experimental,bergen2023large}, giant optical nonlinearities~\cite{morin2022self,makhonin2024nonlinear}, and extremely efficient coupling to microwaves~\cite{pritchett2024giant,gallagher2022microwave}. Very recently, progress has been made on a more applied side with new results in thin-film growth~\cite{delange2023highly}, strong coupling with cavity photons~\cite{orfanakis2022rydberg,makhonin2024nonlinear}, coupling with plasmonic structures~\cite{neubauer2022spectroscopy,ziemkiewicz2022copper} and fabrication methods for the engineering of $\mathrm{Cu_2O}$ micro- and nano-structures~\cite{steinhauer2020rydberg,paul2024local,orfanakis2021quantum} compatible with Rydberg excitons physics.

However, most of these works focused on spectroscopic methods with typically little attention to the time domain. Among the few experimental studies displaying time-resolved measurements~\cite{rogers2022high,bergen2023large,makhonin2024nonlinear,panda2024effective}, none has the resolution to explore Rydberg excitons dynamics within their lifetime, hypothesized from their linewidths to be of order $0.1-100~\si{ps}$. Therefore, the actual lifetime of Rydberg excitons has remained unknown up to now. The situation is similar for their typical coherence time, despite early results on the long-lived (but non-Rydberg) 1S state~\cite{karpinska2005decay,yoshioka2006dark} and later encouraging spectroscopic signatures of coherence within high Rydberg excitons~\cite{grunwald2016signatures}.

\begin{figure*}[t]
\begin{center}
  \includegraphics[width=1\linewidth]{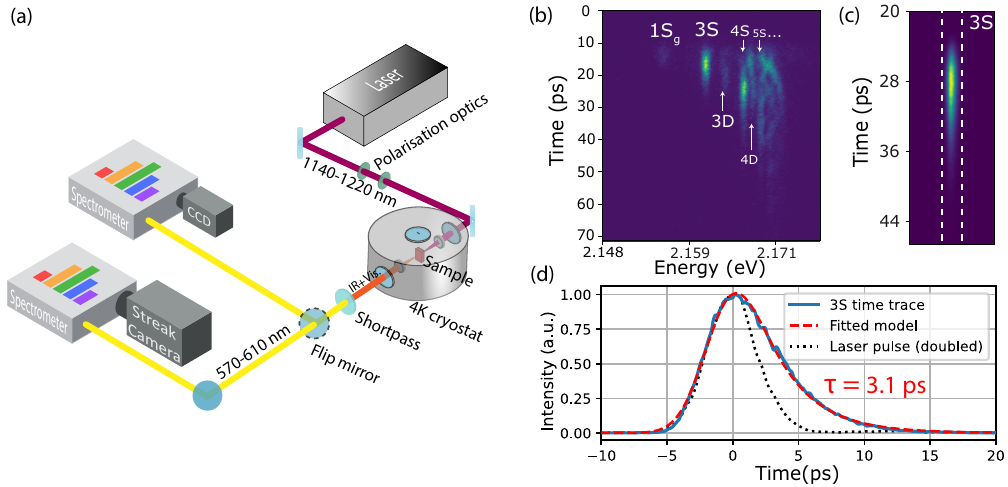}
  \caption{\textit{Experimental protocol}: \textit{(a)} An infrared picosecond laser generates a frequency-doubled signal on the $\mathrm{Cu_2O}$ sample. The signal is collected on a streak camera coupled to a spectrometer to resolve it in both time and energy. A CCD can be used instead of the streak camera to obtain more precise spectra without time resolution. \textit{(b)} Composite image displaying all the Rydberg states in the time-energy domain. \textit{(c)} Example of streak camera image for the 3S state. We integrate along the energy axis (between the white dashed lines) to obtain the time trace. \textit{(d)} 3S time trace for a pump power of 50~$\si{\milli\watt}$. The exciton lifetime is extracted from the fit ($\tau=3.1~\si{\pico\second}$ for the 3S state.)}
  \label{fig1}
  \vspace{-0.5cm}
\end{center}
\end{figure*}

Here we report the first direct lifetime and coherence time measurements (respectively called $T_1$ and $T_2$, following standard terminology) of Rydberg excitons in copper oxide, up to principal quantum number $n = 9$. We rely on a time-resolved version of the so-called second harmonic generation (SHG) spectroscopy~\cite{wang2015giant,Mund2018high} whereby a degenerate two-photon excitation generates, when resonant with an exciton state, a one-photon emission at twice the pump energy. The key ingredient of this work is the high-resolution measurement of the doubled signal dynamics, enabling the observation for the first time of features such as coherent oscillations and the incoherent decay (lifetime) for each of the accessible Rydberg states. Here lifetime refers to the total state lifetime, encompassing both radiative and nonradiative decays. Comparing the lifetime to the (inverse) spectroscopic linewidth of the various states is useful to check for sources of inhomogeneous broadening for $\mathrm{Cu_2O}$ Rydberg excitons. Moreover, we describe the coherent oscillations observed when $n>3$ at low excitation power. We systematically study the effect of the pump power on the dynamics and the coherence time for each state. 


\textbf{Experiment - }We tested two natural $\mathrm{Cu_2O}$ samples from the Tsumeb mine in Namibia with thicknesses $50~\si{\micro\meter}$ and $1~\si{\milli\meter}$. Both are highly polished and cut so that their surface is normal to the [111] crystalline orientation. These samples are cooled in a closed-system helium flow cryostat to $4~\si{\kelvin}$. In the main experiment, a pulsed infrared (IR) laser at half the energy of the Rydberg excitons is sent through the samples and the doubled signal is observed in energy and/or time domain ($0.8~\si{\pico\second}$ absolute resolution, $0.4~\si{\pico\second}$ relative) in transmission mode (figure~\ref{fig1}, see the Supplementary Material~\cite{SM} for details). The samples are oriented such that the light propagates along the [111] crystal axis and we do not apply any electric or magnetic field. In this way, the selection rules dictate that only the even parity states of the S and D series (respective angular momentum $L=0$ and $L=2$) are efficiently excited by the two-photon dipole transition~\cite{Mund2018high}.

As visible in figure~\ref{fig1} (c-d) for the 3S state, the streak camera images are used to produce a time trace for each state. We measure a finite exciton lifetime $\tau=T_1^{(n,L)}$, which we extract for all $(n,L)$ states we can observe. The fit model is the convolution of two distinct contributions: The first is a Gaussian function representing both the temporal width of the laser and the resolution limit of the whole detection system. The second component is an exponential decay from which we extract the lifetime $\tau$ of the state. 

In a second experiment, we independently check the coherence of the signal using an energy-resolved, off-axis Michelson interferometer. The excitation scheme remains the same as in the previous experiment. As presented in figure~S1 of the Supplementary Material~\cite{SM}, the signal is split equally into the two interferometer arms. One arm has a variable delay line ($10~\si{\micro\meter}$ precision) providing a time resolution of about $30~\si{\femto\second}$. In order to get the fringe contrast information in a single camera picture, the two arms are recombined at an angle. That angle is set within the plane perpendicular to the diffraction plane of the spectrometer. This spatial mapping of the fringes removes the need for a sub-wavelength stepping of the mobile mirror (see~\cite{SM} for more details). 

\begin{figure}[t]
\begin{center}
  \includegraphics[width=1\linewidth]{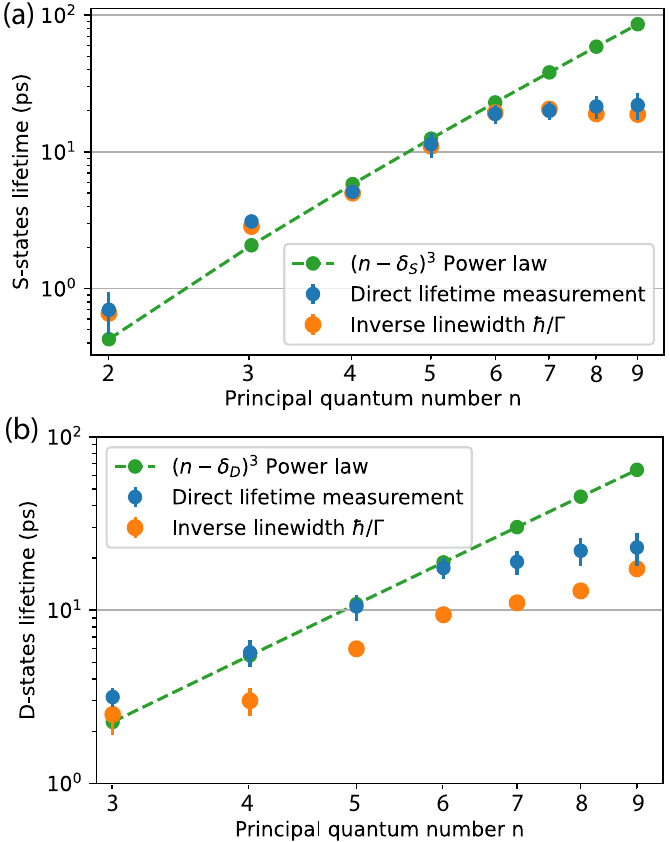}
  \caption{\textit{Lifetimes}: \textit{(a)} Directly measured S-states lifetimes (blue dots), matching the inverse linewidth ($\hbar/\Gamma_\mathrm{nS}$, orange dots). Both are in good agreement with the $(n-\delta_S)^3$ Rydberg scaling (green dots), up to $n=6$. \textit{(b)} Directly measured D-states lifetimes (blue dots). Unlike the S-states, the D-states lifetimes do not match the inverse linewidth ($\hbar/\Gamma_\mathrm{nD}$, orange dots). The agreement with the $(n-\delta_D)^3$ Rydberg scaling (green dots) remains up to $n=6$.}
  \label{fig2}
  \vspace{-0.5cm}
\end{center}
\end{figure}

\begin{table}[t]
    \centering
    \begin{tabular}{|c|c|c|c|} 
        \hline 
        \textbf{S} & \textbf{Energy (eV)} & $\mathbf{\hbar/\Gamma_\mathrm{nS}}$ \textbf{(ps)} & \textbf{Lifetime (ps)}\\ 
        \hline
        1S$^{g}$& 2.15425 & 0.75 $\pm$ 0.1 & 0.74 $\pm$ 0.1 \\  
        \hline  
        2S & 2.13763 & 0.66 $\pm$ 0.04 & 0.7 $\pm$ 0.12  \\  
        \hline 
        3S & 2.16039 & 2.85 $\pm$ 0.02 & 3.1 $\pm$ 0.1 \\  
        \hline 
        4S & 2.16554 & 5.0 $\pm$ 0.05 & 5.1 $\pm$ 0.2\\ 
        \hline 
        5S & 2.16786 & 11.02 $\pm$ 0.05 * & 11.4 $\pm$ 1.2 \\ 
        \hline 
        6S & 2.16922 & 19.37 $\pm$ 0.29 * & 19 $\pm$ 1.5\\ 
        \hline 
        7S & 2.17006 & 20.57 $\pm$ 0.16 * & 20 $\pm$ 1.5\\ 
        \hline
        8S & 2.17053* & 19.37 $\pm$ 0.31 * & 21.5 $\pm$ 2\\ 
        \hline
        9S & 2.17086* & 18.81 $\pm$ 0.4 * & 22 $\pm$ 2.5\\ 
        \hline
        
        \textbf{P} & \textbf{Energy (eV)} & $\mathbf{\hbar/\Gamma_\mathrm{nP}}$ \textbf{(ps)} & \textbf{Lifetime (ps)}\\ 
        \hline
         2P & 2.14732 & 0.33 $\pm$ 0.06 & 0.48 $\pm$ 0.1 \\  
         \hline
         
         \textbf{D} & \textbf{Energy (eV)} & $\mathbf{\hbar/\Gamma_\mathrm{nD}}$ \textbf{(ps)} & \textbf{Lifetime (ps)}\\
         \hline
         $\mathrm{3D_1}$ & 2.16288 & 2.5 $\pm$ 0.3 & 3.2 $\pm$ 0.2 \\ 
         \hline 
         $\mathrm{3D_2}$ & 2.16324 & 2.4 $\pm$ 0.25 & 3.2 $\pm$ 0.2 \\ 
         \hline 
         $\mathrm{4D_1}$ & 2.16642 & 3.0 $\pm$ 0.27 & 5.7 $\pm$ 0.5 \\ 
         \hline 
         $\mathrm{4D_2}$ & 2.16667 & 1.8 $\pm$ 0.1 & 5.7 $\pm$ 0.5 \\ 
         \hline 
         $\mathrm{5D_1}$ & 2.16826 & 3.5 $\pm$ 0.9 & 10.5 $\pm$ 0.9 \\ 
         \hline
         $\mathrm{5D_2}$ & 2.16839 & 5.98 $\pm$ 0.02 * & 10.5 $\pm$ 0.9 \\ 
         \hline
         6D & 2.16945 & 9.4 $\pm$ 0.05 * &  17.5 $\pm$ 1.2 \\ 
         \hline 
         7D & 2.17019* & 11.02 $\pm$ 0.14 * & 19 $\pm$ 1.5 \\ 
         \hline
         8D & 2.17064* & 12.91 $\pm$ 0.16 * & 22 $\pm$ 2 \\ 
         \hline 
         9D & 2.17094* & 17.32 $\pm$ 0.28 * & 23 $\pm$ 2.5 \\ 
         \hline 
    \end{tabular}
    \caption{\textit{Lifetime vs linewidth}: Comparison between the direct lifetime measurement and the inverse linewidth for all measured states (* from ref.~\cite{rogers2022high}).}
    \label{tab:Table 1}
    \vspace{-0.4cm}
\end{table}

\textbf{Results and discussions -} We first discuss the lifetimes $\tau=T_1^{(n,L)}$. The results are presented in Table~\ref{tab:Table 1} for all the states we could access except the yellow 1S state, as it has already been extensively studied in the past~\cite{Sun2001Production,karpinska2005decay,yoshioka2006dark,Mund2019Second}. We present results for the yellow S and D series up to $n=9$, the 1S exciton of the green series and the yellow 2P state which has a weak but allowed dipole-quadrupole transition~\cite{Mund2018high}. Table~\ref{tab:Table 1} also includes the energies of the states as an unambiguous states identifier, as some are known to be mixed~\cite{Schweiner2017even,farenbruch2020rydberg}. The lifetimes of the excited states range from $0.5~\si{\pico\second}$ to over $20~\si{\pico\second}$ and increase systematically with $n$ as predicted by standard Rydberg physics~\cite{heckotter2017scaling}. 

As visible in figure~\ref{fig2} the Rydberg scaling $n^{*3}$, where $n^*=(n-\delta_L)$, works well using the quantum defect values provided by the literature ($\delta_S=0.56$ and $\delta_D=0.08$)~\cite{rogers2022high}, up to $n= 6$. However, the lifetimes remain almost constant for $n\geq 7$ for both the S and the D series. Such plateau has also been observed in their linewidths~\cite{rogers2022high}, which indicates a lifetime-reducing mechanism is present and broadens these states. Albeit less dramatic, this extra broadening has also been observed on the one-photon absorption spectroscopy of the P series~\cite{kazimierczuk2014giant,heckotter2021asymmetric} and explained as an interaction with charged crystal impurities~\cite{kruger2020interaction}. Therefore, we hypothesize that the even series suffer from a similar lifetime-reducing effect, starting at $n=7$ in our samples.

\begin{figure}[t]
\begin{center}
  \includegraphics[width=1\linewidth]{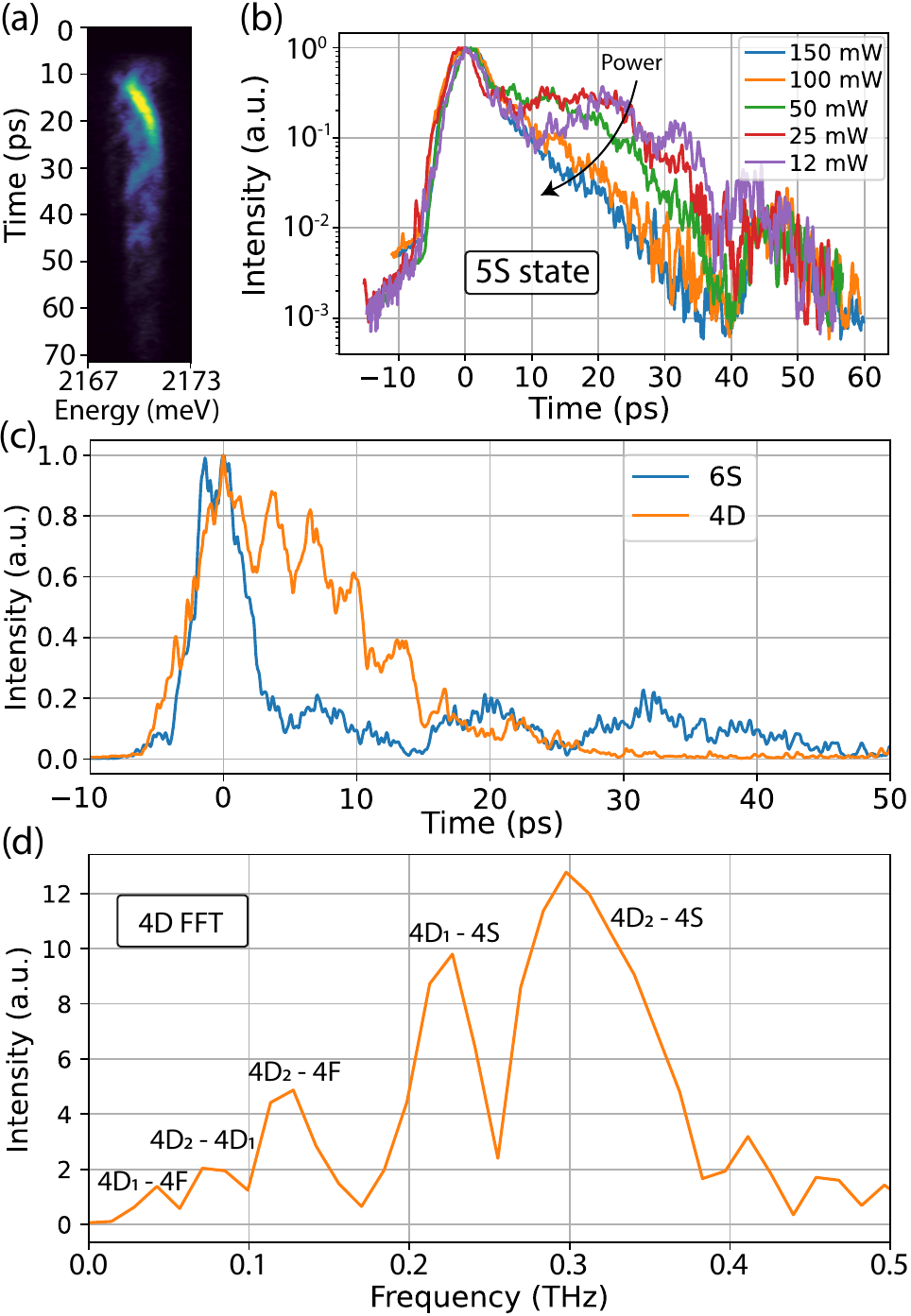}
  \caption{\textit{Oscillations}: \textit{(a)} Example of oscillations typically observed at low pump power (here 50~\si{\milli\watt}) in the time-energy domain on the high Rydberg states region ($n\sim 6\rightarrow\infty$, unresolved). The pump laser is set to 1142.6~\si{\nano\meter}, so that the doubled signal is centered on 2170.21~\si{\milli\electronvolt}. A complex beating pattern involving many states is visible for several tens of picoseconds. \textit{(b)} 5S time traces for various pump powers showing the gradual disappearance of the oscillations. \textit{(c)} Example of two states, 4D and 6S, displaying oscillations with distinct periods (here, 3.3~\si{\pico\second} for 4D and 12~\si{\pico\second} for 6S). \textit{(d)} Numerical Fourier transform of the 4D data revealing several frequencies, each matching the energy separation with a nearby state.}
  \label{fig3}
  \vspace{-0.5cm}
\end{center}
\end{figure}

It is also worth checking the lifetimes against the timescale set by the inverse linewidths $\tau_{min} = \hbar/\Gamma$. Figure~\ref{fig2} and table~\ref{tab:Table 1} present both values for comparison.  Interestingly, all the S states lifetimes follow their inverse linewidths ($\tau\simeq\tau_{min}$), even on the $n\geq 7$ plateau. Therefore, there is no measurable inhomogeneous broadening in the S series (i.e., no purely spectral broadening independent from the lifetime, as opposed to homogeneous broadening where a decreased lifetime broadens the resonance through energy-time relationship). While we expect an homogeneous broadening is present in the data for all states due to phonon scattering, its $n^{*3}$ scaling~\cite{stolz2018interaction} makes it indistinguishable from the usual Rydberg scaling for the raditative lifetime. Our observation has the important implication that the spectroscopic linewidths predict accurately the actual lifetimes of the S series. 

Unlike the S series, the D states lifetimes do not follow their inverse linewidths ($\tau>\tau_{min}$) and live about $2\times$ longer than $\tau_{min}$. While we observe that D states live about as long as or slightly longer than their S counterpart, both our spectra and the literature~\cite{Mund2018high,rogers2022high} measure $n$D states that are significantly broader than $n$S states. Note that each D envelope is not unique, but a non-degenerate multiplet~\cite{Schweiner2017even}. This is unlike the S envelopes which are singlets. As our two-photon excitation selects only D states with $\Gamma_5^{+}$ symmetry, only two distinct D states are present. We respectively label $\mathrm{D_1}$ and $\mathrm{D_2}$ the low and high energy states. While the oscillator strength of the higher energy D state is expected to be $\sim 100\times$ larger than its partner~\cite{Schweiner2017even}, we observe a position-dependent relative signal from both, up to equal signal intensity at some specific sample positions. This allows us to identify the two D states for $n=3-5$. We hypothesize that local strain or impurities can mix the two states, leading to the observed variations. However, even when carefully studying each state separately, their individual linewidths are still significantly broader than predicted by their lifetimes. We chose to stress this on figure~\ref{fig2} (b) by only showing the narrower of the two D states. The origin of this apparent broadening might be the unresolved participation of more states to the measured linewidths. For example, the strain required to resolve the two D states could also mix some non-$\Gamma_5^{+}$ D states. Likewise, the F states lying near the two $\Gamma_5^{+}$ D states could artificially broaden the observed D linewidths. Whatever the cause, our observations indicate that the D series linewidths typically observed with this method do not predict the D states lifetimes. 

Finally, we note that on all states $n\geq4$ the signal dynamics reveals two contributions, as visible in figure~\ref{fig3} (b-c). The first is a strict second harmonic generation process ($\chi^{(2)}$ nonlinearity) enhanced by the presence of exciton resonances, named "SHG" thereafter. The second is a two-photon absorption (TPA, a $\chi^{(3)}$ process) producing exciton states followed by their emission. The presence of both mechanisms was previously recognized for the long-lived 1S state~\cite{Sun2001Production}. We call the latter "Secondary Emission" (SE) to follow earlier terminology in resonant excitation conditions~\cite{haacke1997resonant,garro1999coherent} and to distinguish it from a strict SHG. Unlike SE, SHG is in principle not allowed on the cubic $\mathrm{Cu_2O}$ lattice, but symmetry-breaking factors such as strain and faults allow it weakly~\cite{Mund2019Second}. 


\textit{Oscillations -} We systematically observe time oscillations on the SE signal for states $n\geq 4$ at low pump power. Figure~\ref{fig3} (a) presents a example of raw time-energy data for the highest states (energy just below the gap). The frequency of these oscillations does not depend on the sample thickness and shows a consistent state dependence across both samples. Moreover, in all samples and for all states, the oscillation amplitude decreases with increasing pump power. Therefore, we can rule out any Fabry-Perot effect in the sample or the optics. 

The oscillations completely vanish above some pump power ($P>100~\si{\milli\watt}$ for the $1~\si{\milli\meter}$ sample and $P>250~\si{\milli\watt}$ for the $50~\si{\micro\meter}$ sample). This behavior, exemplified in figure~\ref{fig3} (b) for the 5S state on the thicker sample, is indicative of a coherent process. Such process is expected as the laser linewidth is larger than the energy splitting between neighbor Rydberg states. The decrease of coherence with increasing pump power likely originates from the laser heating the sample, as we recorded a power-dependent rise of the sample holder temperature that was more pronounced for the $1~\si{\milli\meter}$ sample that for the $50~\si{\micro\meter}$ sample. We confirmed this with a temperature study, where rising the temperature while keeping a low pump power showed similar result (see Supplementary Material~\cite{SM}).

\begin{figure}[ht!]
\begin{center}
  \includegraphics[width=1.0\linewidth]{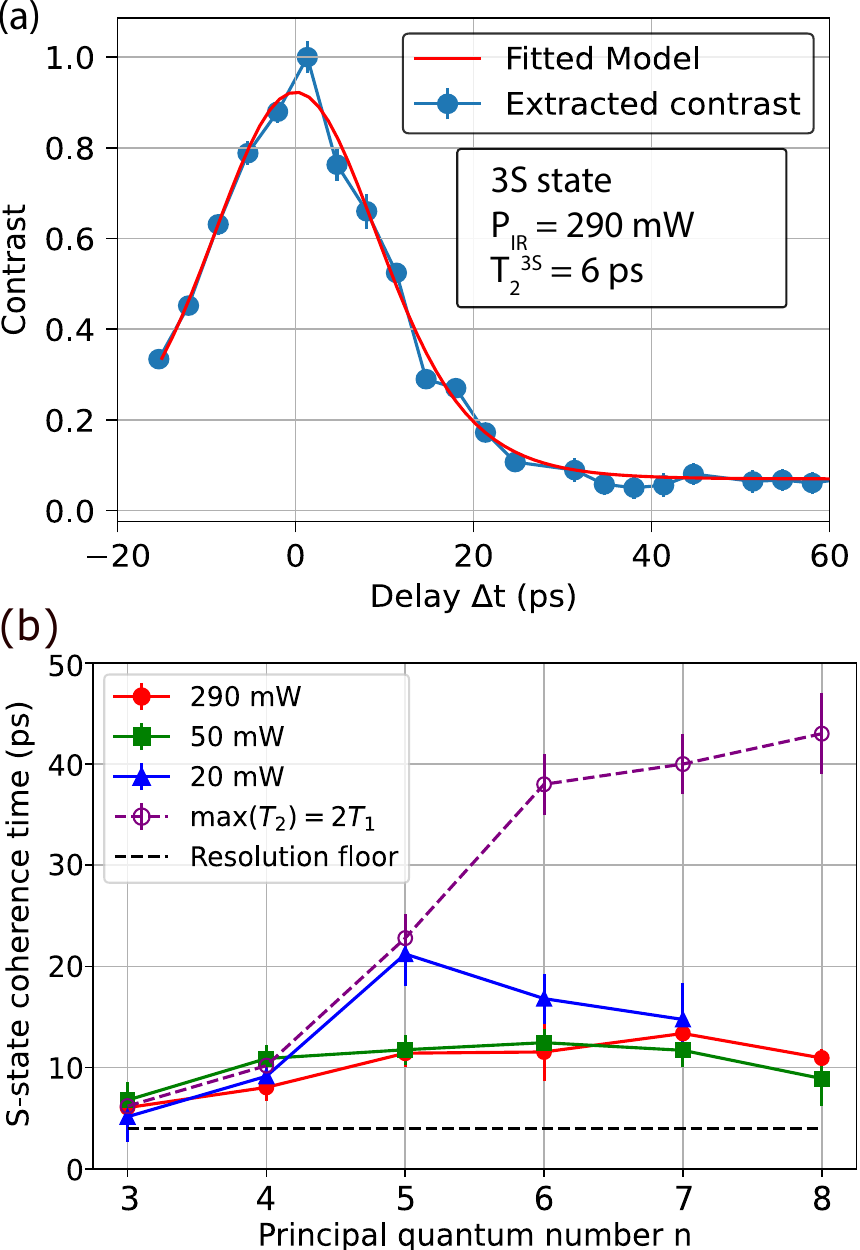}
  \caption{\textit{Coherence times}: \textit{(a)} Typical fringe contrast decay for the 3S state at high pump power. The decay time corresponds to the coherence time $T_2$. \textit{(b)} Coherence time $T_2^{(n,0)}$ as a function of the principal quantum number for the S-states. Three pump powers were used: $290~\si{\milli\watt}$ (red circles), $50~\si{\milli\watt}$ (green squares) and $20~\si{\milli\watt}$ (blue triangles). The theoretical maximum value $\max(T_2^{(n,0)})=2T_1$ (purple rounds and dashed line) and the resolution limit (horizontal dashed line at $\simeq 4~\si{\pico\second}$) are shown for comparison.}
  \label{fig4}
  \vspace{-0.5cm}
\end{center}
\end{figure}

A numerical Fourier analysis of the time traces reveals the oscillations are typically dominated by a main frequency that may be accompanied by a few others of smaller amplitude. These oscillation frequencies are different for each states and decrease with $n$. Figure~\ref{fig3} (c) exemplifies this with the 4D and the 6S states, oscillating at distinct frequencies. Crucially, comparing the energy $h\nu$ associated to each frequency $\nu$ against the energy separation to nearby states shows an excellent match. Table~1 in the Supplementary Material~\cite{SM} lists the main frequency and some of the minor components for $n\geq 4$ states along with their likeliest beating partners, while figure~\ref{fig3} (d) exemplifies this with the 4D Fourier spectrum. We find that for all states, the dominant beating is between S and D states within the same $n$ group. The $n=4-5$ D pairs are especially interesting, as we can clearly distinguish the two beatings that $D_1$ and $D_2$ have their S companion. These beating frequencies are also present in the S dynamics, as expected by reciprocity. We also observe a weak Fourier component at the $D_1$ - $D_2$ energy split, around $70~\si{\giga\hertz}$ for $n=4$ and $40~\si{\giga\hertz}$ for $n=5$.

Importantly, as these measurements are energy-resolved, photons of different energies hit the sensor at different positions and therefore do not overlap. Therefore, we can rule out a purely optical beating between photons originating from different states. This is especially clear on the lowest oscillating states as they are well separated energetically, and therefore spatially on the sensor. Consequently, we infer that these oscillations are a quantum beating between exciton states simultaneously excited by the pump laser.

\textit{Coherence times -} To verify our conclusions we measured the coherence time $T_2^{(n,L)}$ of each state $(n,L)$ for a given pump power with the energy-resolved off-axis interferometer (see figure~S1 and text of the Supplementary Material~\cite{SM}). Figure~\ref{fig4} (a) shows the typical contrast decay as a function of the delay while figure~\ref{fig4} (b) presents the results for the S series. First, we notice that the coherence times are in qualitative agreement with the envelope of the oscillation in the time-resolved measurements. At the lowest power, $T_2$ grows with $n$ and follows very closely the maximum theoretical value of twice the lifetime ($T_2^{\mathrm{max}}=2T_1$) up to $n=5$. Surprisingly, higher states display a shorter coherence time. The decrease of $T_2$ is striking as the power is increased, rapidly stagnating around $12~\si{\pico\second}$ for all states $n\geq 4$. Measuring coherence times longer or similar to the lifetimes is an additional indication that the oscillations result indeed from coherent quantum beats. Together, these observations indicate that the coherence time of Rydberg excitons may be sufficient for more advanced engineering, such as the electromagnetically induced transparency (EIT) schemes that have been proposed~\cite{walther2018giant,walther2020controlling,ziemkiewicz2020electromagnetically} and complement the previous observation of a coherent superposition of neighbor Rydberg exciton states~\cite{grunwald2016signatures}.

\textbf{Conclusion - }In conclusion, this study provides the first direct measurements of lifetimes and coherence times of the even series of Rydberg excitons in $\mathrm{Cu_2O}$ up to a principal quantum number of $n = 9$. Using time-resolved two-photon spectroscopy, we have established that the lifetime of these states follows the expected Rydberg scaling law $n^{*3}$ up to $n = 6$, and confirmed that the plateau previously observed in the linewidth for $n>6$ is also present for the lifetime. Additionally, we observed quantum beats in the time traces of high Rydberg states, which we interpreted as the coherent beating of several excitonic states excited simultaneously. Lastly, we verified their coherence times using an energy-resolved Michelson interferometer. 


Overall, these results suggest that $\mathrm{Cu_2O}$ Rydberg excitons are well-suited for coherent control applications in quantum technologies, laying the groundwork for future exploration of coherent nonlinear effects in this solid-state platform. \\

\begin{acknowledgments}
This work has been supported through the ANR grant ANR-21-CE47-0008 (PIONEEReX), through the EUR grant NanoX ANR-17-EURE-0009 in the framework of the "Programme des Investissements d’Avenir" and through T. Boulier's Junior Professor Chair grant ANR-22-CPJ2-0092-01.
\end{acknowledgments}

\end{document}


\title{Direct measurement of the lifetime and coherence time of $\mathbf{\mathrm{Cu_2O}}$ Rydberg excitons\\-\\Supplementary Informations}

\author{Poulab Chakrabarti}
\affiliation{Universit\'e de Toulouse, INSA-CNRS-UPS, LPCNO, 135 Av. Rangueil, 31077 Toulouse, France}
\author{Kerwan Morin}
\affiliation{Universit\'e de Toulouse, INSA-CNRS-UPS, LPCNO, 135 Av. Rangueil, 31077 Toulouse, France}
\author{Delphine Lagarde}
\affiliation{Universit\'e de Toulouse, INSA-CNRS-UPS, LPCNO, 135 Av. Rangueil, 31077 Toulouse, France}
\author{Xavier Marie}
\affiliation{Universit\'e de Toulouse, INSA-CNRS-UPS, LPCNO, 135 Av. Rangueil, 31077 Toulouse, France}
\affiliation{Institut Universitaire de France, 75231 Paris, France}
\author{Thomas Boulier}
\email{boulier@insa-toulouse.fr}
\affiliation{Universit\'e de Toulouse, INSA-CNRS-UPS, LPCNO, 135 Av. Rangueil, 31077 Toulouse, France}

\date{\today}

\maketitle

\subsection{Experimental details}

\textbf{Streak camera experiment} - As illustrated in figure 1 (a) of the main manuscript, a pulsed Ti:Sa laser pumps an optical parametric oscillator (OPO) to generate the desired wavelength within the range $1140-1220~\si{\nano\meter}$. Each infrared pulse has a temporal duration of $2.44~\si{\pico\second}$ (from a squared  hyperbolic secant fit) and a spectral width of $2.5~\si{\nano\meter}$ (FWHM). As previously reported~\cite{Mund2018high}, the signal intensity is polarization-dependent and we therefore use a circular polarization to maximize the signal. The laser is focused ($20~\si{\micro\meter}$ spot diameter) on the sample. The signal is collected in transmission mode and can be directed towards two different detection systems. The first system consists in a high-end streak camera (Hamamatsu C10910) with sub-picosecond resolution ($\sim 0.8~\si{\pico\second}$ absolute precision, $0.4~\si{\pico\second}$ relative) coupled to a spectrometer. This setup provides both temporal and spectral resolutions (within the time-energy uncertainty relation) and enables direct measurement of the S and D exciton lifetimes, as exemplified in figure 1 (b-c) of the main manuscript. As the various states have vastly different lifetimes and spectral separations with one another, three distinct configurations were used depending on the target state. The first completely removes the spectral resolution by replacing the grating with a mirror, thereby offering the best time resolution (the streak camera resolution). This is used on low $n$ states (typically $n<4$) which have short lifetimes and are spectrally well separated from one another. Here the finite laser linewidth is used to only excite one state at a time. The second configuration uses the spectrometer grating, which vastly reduces the time resolution of the system. To mitigate this, we place beam blockers inside the spectrometer so that only a small central part of the grating is used. This trades off some signal intensity and some spectral resolution for a significant improvement of the time resolution. The active size of the grating can be modulated to change the balance between time and energy resolution. This method typically achieves a resolution of 1.5~$\si{\pico\second}$ with a spectral resolution sufficient to resolve Rydberg states up to $n\simeq 7$. For $n\gtrsim 7$ we infer the states pixel positions on the camera images from our highest resolution spectra, refined by the higher-resolution data found in the literature~\cite{rogers2022high}. The last time-resolved configuration uses a double spectrometer with opposed gratings: Both are set to the same wavelength with opposite orders (+1 and -1) so that their degradation of the time resolution cancels out. Spectral selection is still achieved thanks to the first spectrometer exit slit, that picks out the desired state before sending it to the second spectrometer. This configuration achieves both a good time resolution and a good state selection, at the cost of a significant decrease of the signal intensity. Lastly, the second independent detection system employs a liquid nitrogen-cooled charge-coupled device (CCD) camera coupled to a spectrometer with a shorter period grating to capture higher resolution spectra with no time resolution. The spectral resolution is about $0.2~\si{\milli\electronvolt}$ on the CCD camera and up to $0.6~\si{\milli\electronvolt}$ on the streak camera.\\

\renewcommand{\thefigure}{S\arabic{figure}}
\begin{figure*}[ht!]
\begin{center}
  \includegraphics[width=1\linewidth]{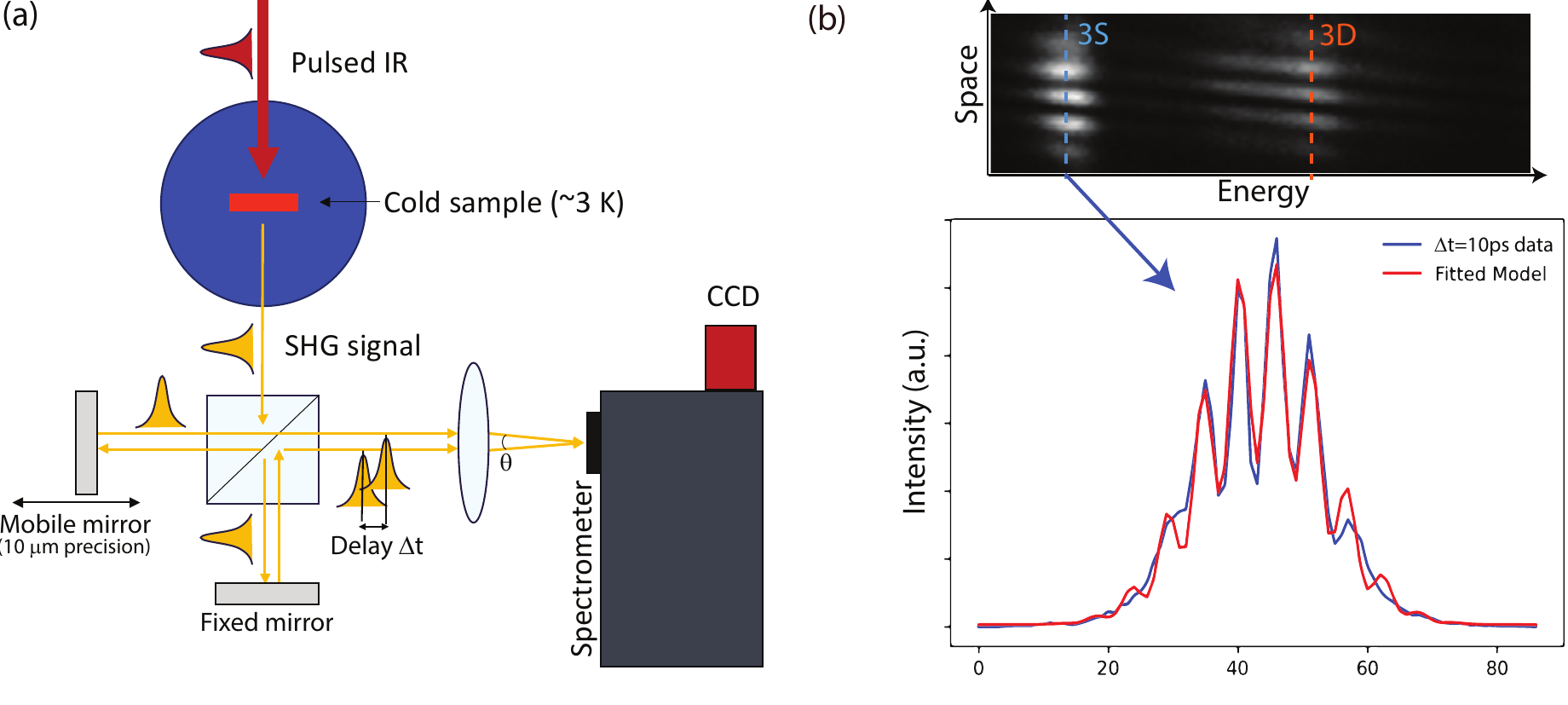}
  \caption{\textit{Coherence}: \textit{(a)} Schematic of the energy-resolved off-axis interferometer. The two interferometer arms arrive on the spectrometer at an angle $\theta$, so that fringes are visible across the non-energy-resolved (vertical) axis of the pictures. \textit{(b)} Example of fringe pattern (top) and single-energy slice (bottom) from which the fringe contrast is extracted.}
  \label{Interferometer}
\end{center}
\end{figure*}

\textbf{Interferometry experiment} - We use an energy-resolved off-axis Michelson interferometer to measure the coherence time of the signal. The excitation scheme remains the same as in the steak camera experiment and a Peltier-cooled CCD camera coupled to a spectrometer is used to resolve in energy the different exciton states. The schematic of the setup is depicted in figure~\ref{Interferometer} (a). A mirror on one of the interferometer arms has a translation stage with a mechanical resolution of $10~\si{\micro\meter}$, so that it can be used as a delay line with about $30~\si{\femto\second}$ time resolution. The signal from the sample is split equally into the interferometer arms and eventually recombines to the spectrometer entrance slit at a small angle in the plane perpendicular to the spectrometer diffraction plane (slit direction). As a result, spatially extended interference fringes are formed, from which the contrast information can be obtained for each delay. This spatial mapping removes the need for a sub-wavelength stepping of the mobile mirror to get the contrast information. Constant-energy cuts in the spectrometer pictures thus give a fringe pattern for each state and for each delay time $\Delta t$, which we model with a Gaussian-modulated cosine function with variable contrast (see figure~\ref{Interferometer} (b)):
\begin{equation}
I(x) = A\exp(-\frac{(x-x_0)^2}{2\sigma^2})\times\frac{1}{2}(C\cos(kx + \phi)+1). 
\end{equation}
The parameter $C$ is the fringe contrast that we seek. It depends on the delay time $\Delta t$ and directly corresponds to the first-order autocorrelation function $g^{(1)}(\Delta t)$. From its decay, we extract the typical timescale during which the signal can interfere with itself; that is to say the coherence time $T_2^{(n,L)}$ of the $(n,L)$ Rydberg state.

\renewcommand{\thefigure}{S\arabic{figure}}
\begin{figure*}[b!]
\begin{center}
  \includegraphics[width=1\linewidth]{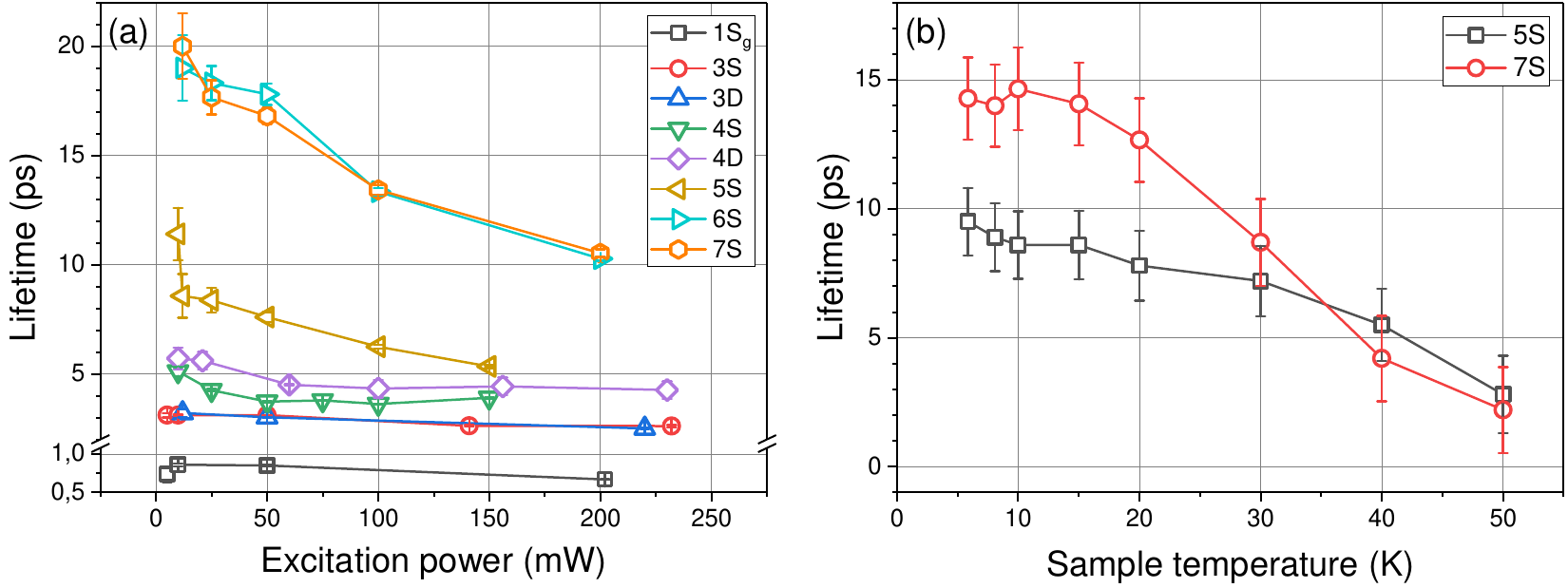}
  \caption{Variation of lifetime for different exciton states with \textit{(a)} the incident laser power and \textit{(b)} the sample temperature.} 
  \label{PowerDep}
\end{center}
\end{figure*}

\subsection{Power and Temperature dependence of the lifetimes}

The exciton lifetime $\tau=T_1^{(n,L)}$ for different $(n,L)$ states are plotted as a function of the average pump power in figure~\ref{PowerDep} (a) for the $1~\si{\milli\meter}$ thick sample. It can be seen that the lifetimes are to some extend power-dependent: they systematically decrease with the pump power for all states. In general, lifetimes are typically reduced by up to $50\%$ between $10~\si{\milli\watt}$ and $200~\si{\milli\watt}$ of IR power, although this seems both sample- and state-dependent. Higher states are more sensitive than lower states. The thinner sample shows smaller lifetimes changes with the pump power, which could indicate a thermal effect is at least partly responsible.

The effect of the temperature on the exciton lifetimes for a fixed pump power of $50~\si{\milli\watt}$ was also investigated for the $1~\si{\milli\meter}$ thick sample. The result is exemplified in panel (b) of figure~\ref{PowerDep} for the 5S and the 7S states. We observe that the lifetimes are decreasing with laser power and that the decrease is more pronounced for the higher n state.

Beyond a certain average pump power, we observe that the sample temperature rises from its lowest value of about $3~\si{\kelvin}$ due to laser heating. This temperature rise was more prominent in the thicker of the two samples. We observe a spectral broadening of all states due to heating, accompanied by the above-mentioned decrease of the lifetimes. We infer from this behavior that the decrease of the lifetimes with the pump power is due to laser heating, and that therefore panels (a) and (b) of figure~\ref{PowerDep} have the same physical origin. However, a more systematic approach would be required to draw definitive conclusions. As the exact origin of these variations is beyond the scope of this article, we report the low-power limit, where the longest lifetimes are observed, in the main manuscript.

\subsection{Table of beating states}

\renewcommand{\thetable}{S\arabic{table}}
\begin{table*}[h!]
\centering 
\begin{tabular}{|p{8mm}||p{51mm}|p{27mm}||p{51mm}|p{29mm}|}
\hline
\textbf{Time trace} & \textbf{Major frequency (energy)} & \textbf{Likely Beating \newline (energy split)} & \textbf{Minor frequencies (energy)} & \textbf{Likely Beating \newline (energy split)} \\ \hline

4S 
& $0.30 \pm 0.01~\si{\tera\hertz}~(1.24 \pm 0.04~\si{\milli\electronvolt})$ 
& 4S-$\mathrm{4D_2}$ ($1.21~\si{\milli\electronvolt}$) 
& $0.21 \pm 0.015~\si{\tera\hertz}~(0.90\pm 0.06~\si{\milli\electronvolt})$ 
& 4S-$\mathrm{4D_1}$ ($0.93~\si{\milli\electronvolt}$) \\ \hline

4D  
& $0.30 \pm 0.02~\si{\tera\hertz}~(1.24 \pm 0.08~\si{\milli\electronvolt})$ 
& $\mathrm{4D_2}$-4S ($1.21~\si{\milli\electronvolt}$) 
& $0.22\pm 0.01~\si{\tera\hertz}~(0.91\pm 0.04~\si{\milli\electronvolt})$ 
\newline $0.12\pm 0.01~\si{\tera\hertz}~(0.50\pm 0.04~\si{\milli\electronvolt})$
\newline $0.07\pm 0.01~\si{\tera\hertz}~(0.29\pm 0.04~\si{\milli\electronvolt})$
\newline $0.05\pm 0.01~\si{\tera\hertz}~(0.21\pm 0.04~\si{\milli\electronvolt})$  
& $\mathrm{4D_1}$-4S ($0.93~\si{\milli\electronvolt}$)
\newline $\mathrm{4D_2}$-4F ($0.50~\si{\milli\electronvolt}$) 
\newline $\mathrm{4D_2}$-$\mathrm{4D_1}$ ($0.29~\si{\milli\electronvolt}$) 
\newline $\mathrm{4D_1}$-4F ($0.21~\si{\milli\electronvolt}$) \\ \hline

5S 
& $0.14 \pm 0.015~\si{\tera\hertz}~(0.58 \pm 0.06~\si{\milli\electronvolt})$ 
& 5S-$\mathrm{5D_2}$ ($0.56~\si{\milli\electronvolt}$)  
& $0.42 \pm 0.015~\si{\tera\hertz}~(1.74 \pm 0.06~\si{\milli\electronvolt})$ 
\newline $0.31 \pm 0.015~\si{\tera\hertz}~(1.30 \pm 0.06~\si{\milli\electronvolt})$
\newline $0.09 \pm 0.015~\si{\tera\hertz}~(0.37 \pm 0.06~\si{\milli\electronvolt})$
& 5S-6D ($1.73~\si{\milli\electronvolt}$)
\newline 5S-6S ($1.28~\si{\milli\electronvolt}$) 
\newline 5S-$\mathrm{5D_1}$ ($0.43~\si{\milli\electronvolt}$)
\\ \hline

5D 
& $0.13 \pm 0.02~\si{\tera\hertz}~(0.54 \pm 0.08~\si{\milli\electronvolt})$
& $\mathrm{5D_2}$-5S ($0.56~\si{\milli\electronvolt}$)
& $0.10 \pm 0.02~\si{\tera\hertz}~(0.41 \pm 0.08~\si{\milli\electronvolt})$ 
\newline $0.04 \pm 0.015~\si{\tera\hertz}~(0.17 \pm 0.06~\si{\milli\electronvolt})$ 
& $\mathrm{5D_1}$-5S ($0.43~\si{\milli\electronvolt}$)
\newline $\mathrm{5D_2}$-$\mathrm{5D_1}$ ($0.13~\si{\milli\electronvolt}$) \\ \hline

6S 
& $0.095 \pm 0.01~\si{\tera\hertz}~(0.39 \pm 0.04~\si{\milli\electronvolt})$ 
& 6S-6D ($0.42~\si{\milli\electronvolt}$) 
& $0.26 \pm 0.01~\si{\tera\hertz}~(1.07 \pm 0.04~\si{\milli\electronvolt})$ 
\newline $0.175 \pm 0.01~\si{\tera\hertz}~(0.72 \pm 0.04~\si{\milli\electronvolt})$ 
& 6S-7D ($1.06~\si{\milli\electronvolt}$*) 
\newline 6S-7S ($0.76~\si{\milli\electronvolt}$*) 
\\ \hline

7S 
& $0.065 \pm 0.01~\si{\tera\hertz}~(0.27 \pm 0.04~\si{\milli\electronvolt})$ 
& 7S-7D ($0.30~\si{\milli\electronvolt}$*) 
& $0.184 \pm 0.01~\si{\tera\hertz}~(0.76 \pm 0.04~\si{\milli\electronvolt})$ 
\newline $0.173 \pm 0.01~\si{\tera\hertz}~(0.72 \pm 0.04~\si{\milli\electronvolt})$ 
\newline $0.11 \pm 0.01~\si{\tera\hertz}~(0.46 \pm 0.04~\si{\milli\electronvolt})$
& 7S-6S ($0.76~\si{\milli\electronvolt}$*) 
\newline 7S-8D ($0.72~\si{\milli\electronvolt}$*) 
\newline 7S-8S ($0.47~\si{\milli\electronvolt}$*) \\ \hline

8S 
& $0.06 \pm 0.01~\si{\tera\hertz}~(0.25 \pm 0.04~\si{\milli\electronvolt})$ 
& 8S-8D ($0.23~\si{\milli\electronvolt}$*) 
& $0.115 \pm 0.01~\si{\tera\hertz}~(0.48 \pm 0.04~\si{\milli\electronvolt})$ 
\newline $0.080 \pm 0.015~\si{\tera\hertz}~(0.33 \pm 0.06~\si{\milli\electronvolt})$ 
\newline $0.040 \pm 0.015~\si{\tera\hertz}~(0.17 \pm 0.06~\si{\milli\electronvolt})$ 
& 8S-7S ($0.48~\si{\milli\electronvolt}$*)
\newline 8S-9S ($0.33~\si{\milli\electronvolt}$*) 
\newline 8S-7D ($0.17~\si{\milli\electronvolt}$*)  \\ \hline

\end{tabular}
\caption{\textit{Quantum beating:} Summary the oscillation frequencies observed in each exciton time traces for $n\geq 4$ states, recorded at low pump power. Typically, a dominant oscillation frequency with maximum amplitude is observed along with a few other frequencies of smaller amplitude. For each frequency, the likeliest states involved in the beating are indicated along with their energy separation. (*) Energy splits for $n\geq 7$ are calculated from ref.~\cite{rogers2022high}.}
\label{tab:Table 2}
\end{table*}


%